\documentclass[preprint,aps,nofootinbib]{revtex4}
\usepackage[dvips]{graphicx}
\usepackage{amssymb}
\usepackage{amsmath}
\usepackage{amsmath, amsthm, amssymb, mathrsfs}
\usepackage{pstricks}

\begin{document}

\title{Three-Tangle for Rank-$3$ Mixed States: mixture of Greenberger-Horne-Zeilinger, 
W and flipped W states}
\author{Eylee Jung, Mi-Ra Hwang,
DaeKil Park}

\affiliation{Department of Physics, Kyungnam University, Masan,
631-701, Korea}

%

\author{Jin-Woo Son}

\affiliation{Department of Mathematics, Kyungnam University, Masan,
631-701, Korea}

\vspace{1.0cm}

\begin{abstract}
Three-tangle for the rank-three mixture composed of Greenberger-Horne-Zeilinger, 
W and flipped W states is analytically calculated. The optimal decompositions in the full
range of parameter space are constructed by making use of the convex-roof extension.
We also provide an analytical technique, which determines whether or not an arbitrary 
rank-$3$ state has vanishing three-tangle. This technique is developed by making use of 
the Bloch sphere $S^8$ of the qutrit system.
The Coffman-Kundu-Wootters inequality is discussed by computing one-tangle and concurrences.
It is shown that the one-tangle is always larger than the sum of squared concurrences and
three-tangle. The physical implication of three-tangle is briefly discussed.
\end{abstract}


\maketitle

Entanglement is a genuine physical resource for the quantum information 
theories\cite{nielsen00}. It is at the heart of the recent much activities on the 
research of quantum computer. Although many new results have been derived recently for the
entanglement of pure states\cite{pure}, entanglement for mixed states is not much 
understood so far compared to the pure states. Since, however, the effect of environment 
generally changes the pure state into the mixed state, it is highly important to 
investigate the entanglement of the mixed states.

Entanglement for the bipartite mixed states, called concurrence, was studied by Hill and
Wootters in Ref.\cite{form2} when the density matrix of the state has two or more 
zero-eigenvalue. Subsequently, Wootters extended the result of Ref.\cite{form2} to the 
arbitrary bipartite mixed states\cite{form3} by making use of the time reversal
operator of the spin-$1/2$ particle appropriately. In addition, the concurrence was used to 
derive the purely tripartite entanglement called residual entanglement or 
three-tangle\cite{tangle1}. For three-qubit pure state 
$|\psi\rangle = \sum_{i,j,k=0}^1 a_{ijk} |ijk\rangle$, the three-tangle $\tau_{3}$
becomes\cite{tangle1}
\begin{equation}
\label{3-tangle-1}
\tau_{3} = 4 |d_1 - 2 d_2 + 4 d_3|,
\end{equation}
where
\begin{eqnarray}
\label{3-tangle-2}
& &d_1 = a^2_{000} a^2_{111} + a^2_{001} a^2_{110} + a^2_{010} a^2_{101} + 
                                                              a^2_{100} a^2_{011
}
                                                              \\   \nonumber
& &d_2 = a_{000} a_{111} a_{011} a_{100} + a_{000} a_{111} a_{101} a_{010} + 
         a_{000} a_{111} a_{110} a_{001}
                                                              \\   \nonumber
& &\hspace{1.0cm} +
         a_{011} a_{100} a_{101} a_{010} + a_{011} a_{100} a_{110} a_{001} + 
         a_{101} a_{010} a_{110} a_{001}
                                                              \\   \nonumber
& &d_3 = a_{000} a_{110} a_{101} a_{011} + a_{111} a_{001} a_{010} a_{100}. 
\end{eqnarray}
The three-tangle is polynomial invariant under the local $SL(2,\mathbb{C})$ 
transformation\cite{ver03,lei04} and exactly coincides with the modulus of a Cayley's 
hyperdeterminant\cite{cay1845,miy03}. For the mixed three-qubit state $\rho$ the three-tangle
is defined by making use of the convex roof construction\cite{benn96,uhlmann99-1} as 
\begin{equation}
\label{3-tangle-3}
\tau_3 (\rho) = \mbox{min} \sum_i p_i \tau_3 (\rho_i),
\end{equation}
where minimum is taken over all possible ensembles of pure states.
The ensemble corresponding to the minimum of $\tau_3$ is called optimal decomposition.

Although the definition of three-tangle for the mixed states is simple as shown in 
Eq.(\ref{3-tangle-3}), it is highly difficult to compute it. This is  
mainly due to the fact that
the construction of the optimal decomposition for the arbitrary state is a formidable task. 
Even for the most simple case of rank-two state still we do not know how to construct the 
optimal decomposition except very rare cases.

Recently, Ref.\cite{tangle2} has shown how to construct the optimal decomposition for the 
rank-$2$ mixture of Greenberger-Horne-Zeilinger(GHZ) and W states:
\begin{equation}
\label{mixture2}
\rho (p) = p |GHZ \rangle \langle GHZ| + (1-p) |W \rangle \langle W|,
\end{equation}
where
\begin{equation}
\label{ghzw}
|GHZ\rangle = \frac{1}{\sqrt{2}} \left( |000\rangle + |111\rangle \right) \hspace{1.0cm}
|W\rangle = \frac{1}{\sqrt{3}} \left( |001\rangle + |010\rangle + |100\rangle \right).
\end{equation}
The optimal decomposition for $\rho(p)$ was constructed with use of the fact that 
$\tau_3 (|GHZ\rangle) = 1$, $\tau_3 (|W\rangle) = 0$ and $\langle GHZ | W \rangle=0$. Once
the optimal decompositions are constructed, it is easy to compute the three-tangle. For 
$\rho(p)$ the three-tangle has three-different expressions depending on the range of $p$ as
following:
\begin{eqnarray}
\label{summary1}
\tau_3 (\rho(p)) = \left\{ \begin{array}{ll}
0 & \hspace{.5cm}   \mbox{for $0 \leq p \leq p_0$}  \\
g_I (p) & \hspace{.5cm}    \mbox{for $p_0 \leq p \leq p_1$}   \\
g_{II} (p) & \hspace{.5cm}    \mbox{for $p_1 \leq p \leq 1$}
                            \end{array}               \right.
\end{eqnarray}
where
\begin{eqnarray}
\label{summary2}
& &g_I (p) = p^2 - \frac{8 \sqrt{6}}{9} \sqrt{p (1-p)^3}  \hspace{1.0cm}
g_{II} (p) = 1 - (1 - p) \left( \frac{3}{2} + \frac{1}{18} \sqrt{465} \right)
                                                                 \\   \nonumber
& &p_0 = \frac{4 \sqrt[3]{2}}{3 + 4 \sqrt[3]{2}} \sim 0.6269  \hspace{1.0cm}
p_1 = \frac{1}{2} + \frac{3}{310} \sqrt{465} \sim 0.7087.
\end{eqnarray}
More recently, this result was extended to the rank-$2$ mixture of generalized GHZ and 
generalized W states in Ref.\cite{tangle3}.

The purpose of this letter is to extend Ref.\cite{tangle2} to the case of rank-$3$ mixed
states. In this paper we would like to analyze the optimal decompositions for the 
mixture of GHZ, W and flipped W states as
\begin{equation}
\label{rank3}
\rho(p,q) = p |GHZ \rangle \langle GHZ| + q |W\rangle \langle W| + 
(1 - p - q) |\tilde{W}\rangle \langle \tilde{W}|,
\end{equation}
where
\begin{equation}
\label{flippedW}
|\tilde{W} \rangle = \frac{1}{\sqrt{3}} \left( |110\rangle + |101\rangle + |011\rangle \right).
\end{equation}
For simplicity, we will define $q$ as 
\begin{equation}
\label{def-q}
q = \frac{1-p}{n},
\end{equation}
where $n$ is positive integer. Before we go further, it is worthwhile noting that 
$\rho(p,q) = \rho(p)$ when $n=1$ and therefore, Eq.(\ref{summary1}) is the three-tangle in 
this case. When $n=\infty$, $\rho(p,q)$ can be constructed from $\rho(p)$ by local-unitary (LU)
transformation $\sigma_x \otimes \sigma_x \otimes \sigma_x$. Since the three-tangle is 
LU-invariant quantity, the three-tangle of $\rho(p,q)$ with $n=\infty$ is again 
Eq.(\ref{summary1}).

Now we start with three-qubit pure state
\begin{equation}
\label{pure1}
|Z (p, q, \varphi_1, \varphi_2) \rangle = \sqrt{p} |GHZ\rangle - e^{i \varphi_1} \sqrt{q}
|W\rangle - e^{i \varphi_2} \sqrt{1 - p - q} |\tilde{W}\rangle
\end{equation}
whose three-tangle is 
\begin{eqnarray}
\label{pure-tangle-1}
& &\tau_3(p, q, \varphi_1, \varphi_2) =    
\bigg| p^2 - 4 p \sqrt{q (1 - p - q)} e^{i(\varphi_1 + \varphi_2)} 
                                                                    \\  \nonumber
& &- \frac{4}{3} q (1-p-q)
e^{2i(\varphi_1 + \varphi_2)} - \frac{8 \sqrt{6}}{9} \sqrt{p q^3} e^{3i \varphi_1} - 
\frac{8 \sqrt{6}}{9} \sqrt{p (1-p-q)^3} e^{3i \varphi_2} \bigg|.
\end{eqnarray}
The state $|Z (p, q, \varphi_1, \varphi_2) \rangle$ has several interesting properties. Firstly,
the mixed state $\rho(p,q)$ in Eq.(\ref{rank3}) can be expressed 
in terms of  $|Z(p,q,\varphi_1,\varphi_2)\rangle$
as following:
\begin{eqnarray}
\label{ensemble1}
& &\rho(p,q) = \frac{1}{3}
\Bigg[ |Z(p,q,0,0)\rangle \langle Z(p,q,0,0)|   \\   \nonumber
& & + 
|Z \left( p,q,\frac{2 \pi}{3}, \frac{4\pi}{3} \right) \rangle
      \langle Z \left(p,q,\frac{2 \pi}{3}, \frac{4\pi}{3}\right)| + 
|Z \left(p,q, \frac{4\pi}{3},\frac{2 \pi}{3} \right)\rangle 
\langle Z \left(p,q, \frac{4\pi}{3},\frac{2 \pi}{3} \right)|
                               \Bigg].
\end{eqnarray}
Secondly, $|Z(p,q,0,0)\rangle$, $|Z \left( p,q,\frac{2 \pi}{3}, \frac{4\pi}{3} \right) \rangle$
and $|Z \left(p,q, \frac{4\pi}{3},\frac{2 \pi}{3} \right)\rangle$ have same three-tangle as
shown from Eq.(\ref{pure-tangle-1}) directly. Thirdly, the numerical calculation shows that
the $p$-dependence of $\tau_3(p,(1-p)/n,\varphi_1,\varphi_2)$ has many zeros depending on 
$\varphi_1$ and $\varphi_2$, but the largest zero $p=p_0$ arises when $\varphi_1=\varphi_2=0$
regardless of $n$. It can be proven rigorously with use of the implicit function theorem.
The $n$-dependence of $p_0$ is given in Table I. Table I indicates that 
when $n$ increases from
$n=2$, $p_0$ approaches to $4 \sqrt[3]{2} / (3 + 4 \sqrt[3]{2}) \sim 0.6269$ from 
$3/4 = 0.75$. This is because of the fact that the three-tangle for $\rho (p,q)$ should be
Eq.(\ref{summary1}) in the $n \rightarrow \infty$ limit.

When $p \leq p_0$, one can construct the optimal decomposition by making use of 
Eq.(\ref{ensemble1}) as following:
\begin{eqnarray}
\label{optimal-least}
& &\rho \left(p, \frac{1-p}{n} \right) = \frac{p}{3 p_0} \Bigg[
|Z\left(p_0,\frac{1-p_0}{n},0,0 \right)\rangle \langle Z\left(p_0,\frac{1-p_0}{n},0,0 \right)|
                                                                       \\   \nonumber
& & \hspace{4.0cm} +
|Z \left(p_0,\frac{1-p_0}{n},\frac{2\pi}{3},\frac{4\pi}{3}\right)\rangle
   \langle Z \left(p_0,\frac{1-p_0}{n},\frac{2\pi}{3},\frac{4\pi}{3}\right) | \\ \nonumber
& & \hspace{4.0cm}   +
|Z \left(p_0,\frac{1-p_0}{n},\frac{4\pi}{3},\frac{2\pi}{3}\right)\rangle
\langle Z \left(p_0,\frac{1-p_0}{n},\frac{4\pi}{3},\frac{2\pi}{3}\right)| \Bigg]
                                                                         \\  \nonumber
& & \hspace{3.0cm}   + \frac{p_0 - p}{n p_0} |W\rangle \langle W| + 
\frac{(n-1) (p_0 - p)}{n p_0} |\tilde{W}\rangle \langle \tilde{W}|.
\end{eqnarray}
Thus, we have vanishing three-tangle in this region:
\begin{equation}
\label{tangle-least}
\tau_{3} \left[\rho \left(p, \frac{1-p}{n} \right) \right] = 0 \hspace{1.0cm}
\mbox{for $p \leq p_0$}.
\end{equation}

Now, we consider $p_0 \leq p \leq 1$ region. When $p = p_0$, Eq.(\ref{optimal-least}) implies
that the optimal decomposition consists of three pure states
$| Z\left(p_0, \frac{1 - p_0}{n}, 0, 0\right)\rangle$, 
$| Z\left(p_0, \frac{1 - p_0}{n}, \frac{2\pi}{3}, \frac{4\pi}{3} \right)\rangle$, and 
$| Z\left(p_0, \frac{1 - p_0}{n}, \frac{4\pi}{3}, \frac{2\pi}{3} \right)\rangle$ with same 
probability. This fact together with Eq.(\ref{ensemble1}) strongly suggests that the 
optimal decomposition at $p_0 \leq p$ is described by Eq.(\ref{ensemble1}). As will be 
shown below, however, this is not true in the full range of $p_0 \leq p \leq 1$.

The optimal decomposition (\ref{ensemble1}) gives the three-tangle to $\rho(p,q)$ in
a form
\begin{equation}
\label{former-1}
\alpha_I (p) = p^2 - \frac{4 \sqrt{n-1}}{n} p (1-p) - \frac{4 (n-1)}{3 n^2} (1-p)^2
- \frac{8 \sqrt{6 n} \left[1 + (n-1)^{3/2}\right]}{9 n^2} \sqrt{p (1 - p)^3}.
\end{equation}
Since the three-tangle for mixed state is defined as a convex roof, $\alpha_I (p)$ should
be convex function if it is a correct three-tangle in the range of $p_0 \leq p \leq 1$. In 
order to check this we compute $d^2 \alpha_I / dp^2$, which is 
\begin{equation}
\label{former-2}
\frac{d^2 \alpha_I (p)}{dp^2} = 
\frac{2}{9 n^2} \left[ \left\{ 9 n^2 + 36 n \sqrt{n-1} - 12 (n-1) \right\} - 
\sqrt{6 n} \left\{1 + (n-1)^{3/2} \right\}
\frac{8 p^2 - 4 p - 1}{\sqrt{p^3 (1-p)}} \right].
\end{equation}
Using Eq.(\ref{former-2}) one can show that $d^2 \alpha_I (p) / dp^2 \leq 0$ when 
$p_* \leq p \leq 1$. The $n$-dependence of $p_*$ is given in Table I. Thus, we need to
convexify $\alpha_I (p)$ in the region $p_1 \leq p \leq 1$, where $p_1 \leq p_*$. The constant
$p_1$ will be determined shortly.

\begin{figure}[ht!]
\begin{center}
\includegraphics[height=5.0cm]{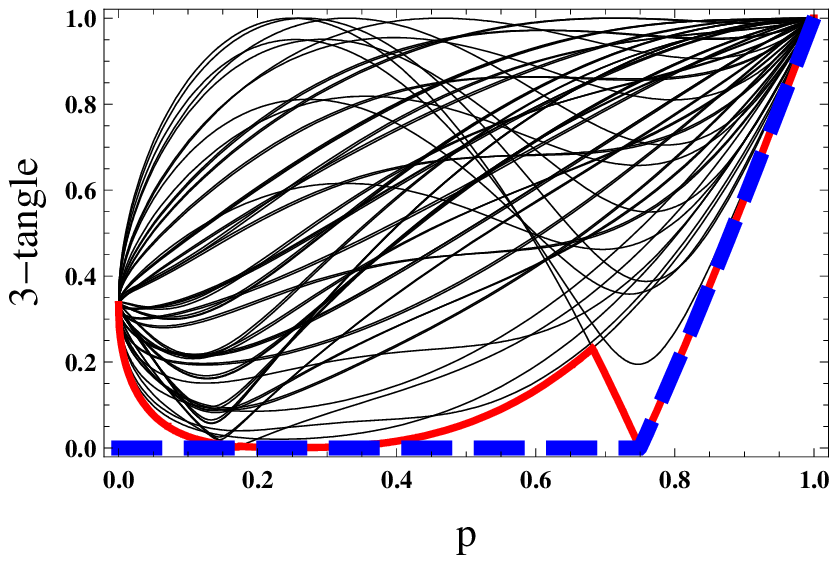}
\includegraphics[height=5.0cm]{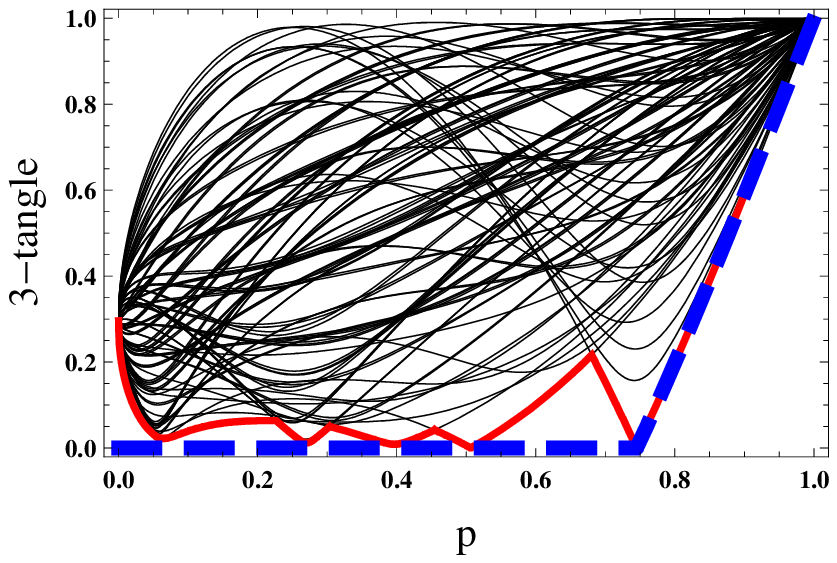}
\includegraphics[height=5.0cm]{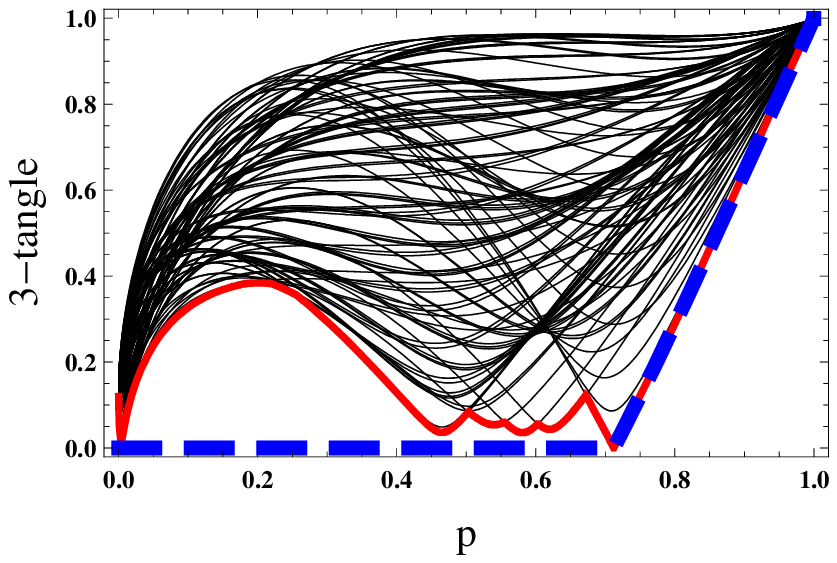}
\caption[fig1]{(color online) 
The plot of $p$-dependence of the Eq.(\ref{pure-tangle-1}) for various
$\varphi_1$ and $\varphi_2$. We have chosen $\varphi_1$ and $\varphi_2$ from $0$ to $2\pi$
as an interval $0.3$. The three figures correspond to $n=2$ (Fig. 1a), $n=3$ (Fig. 1b) and 
$n=10$ (Fig. 1c) respectively. The minimum curve 
is plotted as a thick solid line in each figure. These
figures indicate that the three-tangle in Eq.(\ref{final-1}) (plotted as dashed line in each
figure) is a convex hull of the thick solid line.}
\end{center}
\end{figure}

For large $p$ region one can construct the optimal decomposition as following:
\begin{eqnarray}
\label{latter-1}
\rho(p, q)&=& p |GHZ\rangle \langle GHZ| + \frac{1-p}{n} |W\rangle \langle W| + 
                \frac{(n-1) (1-p)}{n} |\tilde{W}\rangle \langle \tilde{W}|
                                                                   \\   \nonumber
&=& p|GHZ\rangle \langle GHZ| + 
\frac{1-p}{1-p_1} \bigg[ -p_1|GHZ\rangle \langle GHZ| + p_1 |GHZ\rangle \langle GHZ| 
                                                                   \\   \nonumber
& & \hspace{1.0cm} +
                         \frac{1-p_1}{n} |W\rangle \langle W| + 
                      \frac{(n-1) (1 - p_1)}{n} |\tilde{W}\rangle \langle \tilde{W}| \bigg]
                                                                    \\   \nonumber
&=& \frac{p-p_1}{1-p_1} |GHZ\rangle \langle GHZ|
+
    \frac{1-p}{3(1 - p_1)} \Bigg[ |Z\left(p_1, \frac{1-p_1}{n},0,0\right)\rangle
\langle Z\left(p_1, \frac{1-p_1}{n},0,0\right)| 
                                                                      \\   \nonumber
& & \hspace{3.0cm}+ 
|Z\left(p_1, \frac{1-p_1}{n},\frac{2\pi}{3}, \frac{4\pi}{3}\right)\rangle
\langle Z\left(p_1, \frac{1-p_1}{n},\frac{2\pi}{3}, \frac{4\pi}{3}\right)| 
                                                                        \\   \nonumber
& & \hspace{3.0cm}+ 
|Z\left(p_1, \frac{1-p_1}{n},\frac{4\pi}{3}, \frac{2\pi}{3}\right)\rangle
\langle Z\left(p_1, \frac{1-p_1}{n},\frac{4\pi}{3}, \frac{2\pi}{3}\right)|
                                                                          \Bigg]
\end{eqnarray}
which gives the three-tangle in a form
\begin{equation}
\label{latter-2}
\alpha_{II} (p) = \frac{p-p_1}{1-p_1} + \frac{1-p}{1-p_1} \alpha_I (p_1).
\end{equation}
Note that $d^2 \alpha_{II} (p) / dp^2 = 0$. Thus, $\alpha_{II}(p)$ does not violate the convex 
constraint of the three-tangle in the large $p$ region. The parameter $p_1$ is determined by 
minimizing $\alpha_{II}(p)$, {\it i.e.} $\partial \alpha_{II} / \partial p_1 = 0$, which gives
\begin{equation}
\label{latter-3}
\frac{4 \sqrt{6 n} \left[1 + (n-1)^{3/2} \right]}{9 n^2}
\frac{2 p_1 - 1}{\sqrt{p_1 (1 - p_1)}} = 1 + \frac{4 \sqrt{n-1}}{n} - \frac{4 (n-1)}{3 n^2}.
\end{equation}

\begin{center}
\begin{tabular}{c|cccccc} \hline
$n$ & $1$ & $2$ & $3$ & $10$ & $100$ & $1000$    \\   \hline  \hline
$p_0$ & $0.6269$ & $0.75$ & $0.7452$ & $0.712$ & $0.6604$ & $0.6382$   \\  \hline
$p_1$ & $0.7087$ & $0.9330$ & $0.9250$ & $0.8667$ & $0.7710$ & $0.7298$   \\
                                                                                   \hline
$p_*$ & $0.8257$ & $0.9618$ & $0.9572$ & $0.9230$ & $0.8650$ & $0.8391$  \\
                                                                                     \hline
\end{tabular}

\vspace{0.1cm}
Table I: The $n$-dependence of $p_0$, $p_1$ and $p_*$.
\end{center}
\vspace{0.5cm}
The $n$-dependence of $p_1$ is given in Table I. As expected $p_1$ is between $p_0$ and $p_*$.
When $n$ increases from $n=2$, $p_1$ decreases from $(2 + \sqrt{3}) / 4 \sim 0.933$ to 
$1/2 + 3 \sqrt{465} / 310 \sim 0.709$.

\begin{figure}[ht!]
\begin{center}
\includegraphics[height=10.0cm]{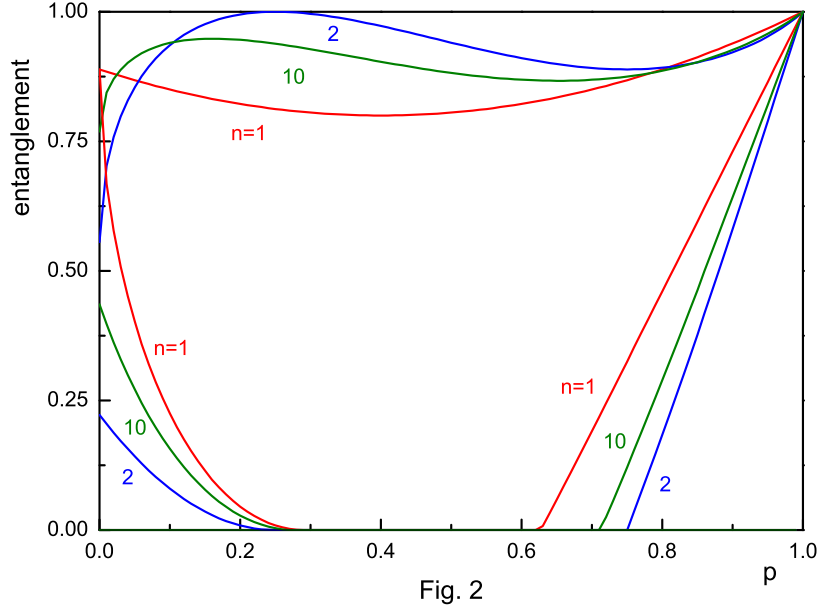}
\caption[fig2]{(color online) The $p$-dependence of one-tangle (upper solidlines), 
sum of squared 
concurrences (left solid lines) and three-tangle (right solid lines) for $n=1$, 
$2$ and $10$. This figure clearly indicates that not only
CKW inequality (\ref{ckw2}) but also (\ref{ckw3}) hold for all integer $n$.}  
\end{center}
\end{figure}

In summary, the three-tangle for $\rho(p,q)$ is 
\begin{eqnarray}
\label{final-1}
\tau_3 (\rho(p,q)) = \left\{ \begin{array}{ll}
0 & \hspace{.5cm}   \mbox{for $0 \leq p \leq p_0$}  \\
\alpha_I (p) & \hspace{.5cm}    \mbox{for $p_0 \leq p \leq p_1$}   \\
\alpha_{II} (p) & \hspace{.5cm}    \mbox{for $p_1 \leq p \leq 1$}
                            \end{array}               \right.
\end{eqnarray}
and the corresponding optimal decompositions are (\ref{optimal-least}), (\ref{ensemble1}), 
and (\ref{latter-1}) respectively.
In order to show that Eq.(\ref{final-1}) is genuine optimal, we plotted the $p$-dependence 
of the three-tangles (\ref{pure-tangle-1}) for various $\varphi_1$ and $\varphi_2$ when
$n=2$ (Fig. 1a), $n=3$ (Fig. 1b) and $n=10$ (Fig. 1c). These curves have been referred as
the characteristic curves\cite{oster07}. As Ref.\cite{oster07} indicated, the three-tangle is 
a convex hull of the minimum of the characteristic curves (thick solid lines in the figure). 
Fig. 1 indicates that the three-tangles (\ref{final-1}) plotted as dashed lines are the 
convex characteristic curves, which implies that Eq.(\ref{final-1}) is really optimal.

The above analysis can be applied to provide an analytical technique which decides whether 
or not an arbitrary rank-$3$ state has vanishing three-tangle. First we correspond
our states to the qutrit states with 
\begin{eqnarray}
\label{bloch-1}
|GHZ\rangle = \left( \begin{array}{c} 1 \\  0  \\ 0 \end{array} \right) \hspace{1.0cm}
|W\rangle = \left( \begin{array}{c} 0 \\  1  \\ 0 \end{array} \right) \hspace{1.0cm}
|\tilde{W}\rangle = \left( \begin{array}{c} 0 \\  0  \\ 1 \end{array} \right).
\end{eqnarray}
It is well-known\cite{qutrit-1} that the density matrix of the arbitrary qutrit state
can be represented by $\rho = (1/3) (I + \sqrt{3} \vec{n} \cdot \vec{\lambda})$, where
$\vec{n}$ is $8$-dimensional unit vector and $\lambda_i \hspace{.2cm} (i=1, \cdots, 8)$ are 
Gell-Mann matrices. Thus the points on the $S^8$ correspond to pure qutrit states while
the interior points denote the mixed states\footnote{Unlike qubit system not all points in
$S^8$ do correspond to the qutrit states due to the condition of star product\cite{qutrit-1}}.
Then, one can show straightforwardly that the pure states with vanishing three-tangle 
correspond to
\begin{eqnarray}
\label{bloch-2}
& &|W\rangle \rightarrow \left(0, 0, -\frac{\sqrt{3}}{2}, 0, 0, 0, 0, \frac{1}{2} \right)
                                                                          \\   \nonumber
& &|\tilde{W}\rangle \rightarrow \left(0, 0, 0, 0, 0, 0, 0, -1 \right)   \\  \nonumber
& &| Z\left(p_0, \frac{1 - p_0}{n}, 0, 0 \right) \rangle \rightarrow
\left(-\sqrt{3} \xi_1, 0, \eta_1, -\sqrt{3} \xi_2, 0, \sqrt{3} \xi_3, 0, \eta_2 \right)
                                                                          \\   \nonumber
& &| Z\left(p_0, \frac{1 - p_0}{n}, \frac{2\pi}{3}, \frac{4\pi}{3} \right) \rangle \rightarrow
\left(\frac{\sqrt{3}}{2} \xi_1, -\frac{3}{2} \xi_1, \eta_1, \frac{\sqrt{3}}{2} \xi_2, 
\frac{3}{2} \xi_2, -\frac{\sqrt{3}}{2} \xi_3, \frac{3}{2} \xi_3, \eta_2 \right)
                                                                          \\   \nonumber
& &| Z\left(p_0, \frac{1 - p_0}{n}, \frac{4\pi}{3}, \frac{2\pi}{3} \right) \rangle \rightarrow
\left(\frac{\sqrt{3}}{2} \xi_1, \frac{3}{2} \xi_1, \eta_1, \frac{\sqrt{3}}{2} \xi_2, 
-\frac{3}{2} \xi_2, -\frac{\sqrt{3}}{2} \xi_3, -\frac{3}{2} \xi_3, \eta_2 \right),
\end{eqnarray}
where $\xi_1 = \sqrt{p_0 (1-p_0) / n}$, $\xi_2 = \sqrt{n-1} \xi_1$,
$\xi_3 = \sqrt{n-1} (1-p_0) / n$, $\eta_1 = (\sqrt{3}/2) (1 - (n+1) (1 - p_0)/n)$ and 
$\eta_2 = (1/2) (1 - 3 (n-1) (1-p_0) / n)$. Thus these five points form a hyper-polyhedron in
$8$-dimensional space. Then all rank-$3$ quantum states corresponding to the points in this
hyper-polyhedron have vanishing three-tangle. 

Now we would like to consider the Coffman-Kundu-Wootters(CKW) relation\cite{tangle1},
which is 
\begin{equation}
\label{ckw1}
4 \mbox{det} \rho_A = {\cal C}_{AB}^2 + {\cal C}_{AC}^2 + \tau_3 (\psi)
\end{equation}
for three-qubit pure state $|\psi\rangle$. In Eq.(\ref{ckw1}) ${\cal C}_{AB}$ and 
${\cal C}_{AC}$ are the concurrences for the corresponding reduced states. Eq.(\ref{ckw1})
indicates that the entanglement of qubit $A$ is originated from both bipartite and 
tripartite contributions. For mixed state Ref.\cite{tangle1} has shown
\begin{equation}
\label{ckw2}
4 \min \left[ \mbox{det} (\rho_A) \right] \geq {\cal C}_{AB}^2 + {\cal C}_{AC}^2,
\end{equation}
where minimum of one-tangle is taken over all possible decompositions of $\rho$. 
In Ref.\cite{tangle2} the CKW inequality (\ref{ckw2}) has been examined for the mixture of 
GHZ and W states. For this case it was shown that the one-tangle is always larger than the 
sum of squared concurrences and three-tangle.

Now, we would like to check the CKW inequality for $\rho(p, q)$ in Eq.(\ref{rank3}) with
$q = (1-p) / n$. In this case one can compute the minimum one-tangle directly, whose
expression is 
\begin{eqnarray}
\label{one-tangle1}
& &4 \min \left[ \mbox{det} \rho_A \right]   
= \frac{1}{9}
\bigg[ (8 - 4 p - 12 q + 5 p^2 + 12 q^2 + 12 p q)               \\  \nonumber 
& &  \hspace{3.0cm} 
+ 4 \sqrt{p q (1 - p - q)}
      \left(2 \sqrt{6 q} + 2 \sqrt{6 (1 - p - q)} - 3 \sqrt{p} \right) \bigg].
\end{eqnarray}
Also it is straightforward to compute the sum of squared concurrences, which is 
\begin{equation}
\label{two-tangle}
{\cal C}_{AB}^2 + {\cal C}_{AC}^2 = 2 \left( \max \left[0,
\frac{2}{3} (1 - p) - \frac{1}{3} \sqrt{(3 p + 2 q) (2 + p - 2 q)} \right] \right)^2.
\end{equation}
The one-tangle(upper solid lines), ${\cal C}_{AB}^2 + {\cal C}_{AC}^2$(left solid lines), and 
three-tangle(right solid lines) are plotted in Fig. 1 for $n=1$, $n=2$ and $n=10$.
This figure indicates that all quantities approach to their
corresponding $n=1$ quantity when $n$ increases from $n=2$. This is consistent with the
fact that $\rho(p, q)$ with $n=1$ is LU-equivalent to $\rho(p, q)$ with $n=\infty$. The 
inequality
\begin{equation}
\label{ckw3}
4 \min \left[ \mbox{det} (\rho_A) \right] \geq {\cal C}_{AB}^2 + {\cal C}_{AC}^2 + \tau_3
\end{equation}
holds for all $n$. In the region $p_C \leq p \leq p_0$, where
\begin{equation}
\label{p-C}
p_C = \frac{(7 n^2 - 4 n + 4) - 3 n \sqrt{5 n^2 - 4 n + 4}}{(n-2)^2},
\end{equation}
both ${\cal C}_{AB}^2 + {\cal C}_{AC}^2$ and $\tau_3$ vanish while there is quite substantial 
one-tangle. Its interpretation is given in Ref.\cite{tangle2} from the mathematical point
of view. However, its physical meaning is still unclear at least for us. In the region
$p \geq p_C$ and $p \leq p_0$ the entanglement of the qubit $A$ mainly stems from the 
bipartite and tripartite correlations, respectively.

One may wonder why we do not take $q = \alpha (1-p)$ with real number $0 \leq \alpha \leq 1$.
For this case, however, it is unclear whether or not the $p$-dependence of 
$\tau_3 (p, q, \varphi_1, \varphi_2)$ in Eq.(\ref{pure-tangle-1}) has maximum zero at 
$\varphi_1 = \varphi_2 = 0$ regardless of $\alpha$. If this is correct, our result can be 
easily extended to the case of $q = \alpha (1-p)$ by changing $n \rightarrow 1/\alpha$.

There are many rank-$3$ mixed states whose three-tangles may exhibit interesting behavior. 
For example, let us consider the state
\begin{equation}
\label{conclu-1}
\pi(p,n) = p |GHZ,+\rangle \langle GHZ,+| + \frac{1-p}{n} |W\rangle \langle W| + 
\frac{(n-1) (1-p)}{n} |GHZ,-\rangle \langle GHZ,-|,
\end{equation}
where $|GHZ,\pm \rangle = (1/\sqrt{2}) (|000\rangle \pm |111\rangle)$. 
Unlike $\rho (p, n)$ discussed in the present paper $\pi (p,1)$ is not LU-equivalent with
$\pi (p, \infty)$. 
When $n=1$, 
$\pi(p,1)$ is identical with $\rho(p,q)$ with $n=1$. When $n=\infty$, the three-tangle of 
$\pi(p,\infty)$ can be calculated by similar method and the result is $(2p - 1)^2$. If 
$n$ increases from $n=2$, the three-tangle should move to $(2p - 1)^2$ from Eq.(\ref{summary1})
smoothly. The particular point $p=1/2$ may 
play a role as a fixed point. It is interesting to examine this behavior by deriving the 
optimal decomposition of $\pi(p,n)$ in the full range of $p$ and $n$.

Of course, it is extremely important if we develop a calculational technique, which enables
us to compute the three-tangle for the arbitrary mixed states.
In order to explore this issue we should develop a technique first, which enables us to
compute the three-tangle for the arbitrary rank-two mixed states as Hill and Wootters 
did in the concurrence calculation in Ref.\cite{form2}. For the case of concurrence, however,
Hill and Wootters exploited fully the magic properties of the magic basis 
$\{|e_i\rangle, i = 1,\cdots ,4 \}$. In this basis the concurrence for the two-qubit state
$|\psi\rangle$ can be expressed as $|\sum \alpha_i^2|$, where 
$|\psi\rangle = \sum_i \alpha_i |e_i\rangle$. Then this property and usual convexification
technique make it possible to compute the concurrence for the arbitrary rank-two bipartite
mixed states. Such a basis, however, is not found in the three-qubit system so far. 
Furthermore, we do not know whether or not such a basis exists in the higher-qubit system.
Thus it is very difficult problem to go further this issue.

From the aspect of physics it is also of interest to investigate the physical role of the 
three-tangle. As shown in Ref.\cite{08-mixed} the two-qubit mixed-state entanglement 
provides an information on the fidelity in the bipartite teleportation through noisy
channels. Since the three-tangle is purely tripartite entanglement, it may give
certain information in the scheme of quantum copy machine or three-party quantum
teleportation\cite{karl98}. 
It seems to be interesting to explore the physical role of the three-tangle in the 
particular real tasks.

{\bf Acknowledgement}: 
This work was supported by the Kyungnam University
Foundation Grant, 2008.


\begin{thebibliography}{99}
\bibitem{nielsen00} M. A. Nielsen and I. L. Chuang, {\it Quantum Computation and 
Quantum Information} (Cambridge University Press, Cambridge, England, 2000).
\bibitem{pure} T. C. Wei and P. M. Goldbart, {\it Geometric measure of entanglement and 
applications to bipartite and multipartite quantum states}, Phys. Rev. {\bf A68} (2003)
042307 [quant-ph/0307219]; E. Jung, M. R. Hwang, H. Kim, M. S. Kim, D. K. Park, J. W. Son
and S. Tamaryan, {\it Reduced State Uniquely Defines Groverian Measure of Original Pure
State}, Phys. Rev. {\bf A77} (2008) 062317
[arXiv:0709.4292 (quant-ph)]; L. Tamaryan, DaeKil K. Park and S. Tamaryan,
{\it Analytic Expressions for Geometric Measure of Three Qubit
States}, Phys. Rev. {\bf A 77} (2008) 022325,
[arXiv:0710.0571 (quant-ph)]; L. Tamaryan, DaeKil Park, Jin-Woo Son, S. Tamaryan, {\it Geometric
Measure of Entanglement and Shared Quantum States}, Phys. Rev. {\bf A78} (2008) 032304,
[arXiv:0803.1040 (quant-ph)]; E. Jung, Mi-Ra Hwang, DaeKil Park, L. Tamaryan and S. Tamaryan,
{\it Three-Qubit Groverian Measure}, Quant. Inf. Comp. {\bf 8} (2008) 0925
[arXiv:0803.3311 (quant-ph)].
\bibitem{form2} S. Hill and W. K. Wootters, {\it Entanglement of a Pair of Quantum Bits},
Phys. Rev. Lett. {\bf 78} (1997) 5022 [quant-ph/9703041].
\bibitem{form3} W. K. Wootters, {\it Entanglement of Formation of an Arbitrary State
of Two Qubits}, Phys. Rev. Lett. {\bf 80} (1998) 2245 [quant-ph/9709029].
\bibitem{tangle1} V. Coffman, J. Kundu and W. K. Wootters, {\it Distributed entanglement},
Phys. Rev. {\bf A61} (2000) 052306 [quant-ph/9907047].
\bibitem{ver03} F. Verstraete, J. Dehaene and B. D. Moor, {\it Normal forms and entanglement
measures for multipartite quantum states}, Phys. Rev. {\bf A68} (2003) 012103
[quant-ph/0105090].
\bibitem{lei04} M. S. Leifer, N. Linden and A. Winter, {\it Measuring polynomial invariants 
of multiparty quantum states}, Phys. Rev. {\bf A69} (2004) 052304 [quant-ph/0308008].
\bibitem{cay1845} A. Cayley, {\it On the Theory of Linear Transformations}, Cambridge Math. 
J. {\bf 4} (1845) 193.
\bibitem{miy03} A. Miyake, {\it Classification of multipartite entangled states 
by multidimensional determinants}, Phys. Rev. {\bf A67} (2003) 012108 [quant-ph/0206111].
\bibitem{benn96} C. H. Bennett, D. P. DiVincenzo, J. A. Smokin and W. K. Wootters,
{\it Mixed-state entanglement and quantum error correction}, Phys. Rev. {\bf A54}
(1996) 3824 [quant-ph/9604024].
\bibitem{uhlmann99-1} A. Uhlmann, {\it Fidelity and concurrence of conjugate states},
Phys. Rev. {\bf A62} (2000) 032307 [quant-ph/9909060].
\bibitem{tangle2} R. Lohmayer, A. Osterloh, J. Siewert and A. Uhlmann, {\it Entangled
Three-Qubit States without Concurrence and Three-Tangle}, Phys. Rev. Lett. {\bf 97}
(2006) 260502 [quant-ph/0606071].
\bibitem{tangle3} C. Eltschka, A. Osterloh, J. Siewert and A. Uhlmann, {\it Three-tangle
for mixtures of generalized GHZ and generalized W states}, arXiv:0711.4477 (quant-ph).
\bibitem{oster07} A. Osterloh, J. Siewert and A. Uhlmann. {\it Tangles of superpositions and
the convex-roof extension}, arXiv:0710.5909 [quant-ph].
\bibitem{qutrit-1} C. M. Caves and G. J. Milburn, {\it Qutrit Entanglement}, 
quant-ph/9910001.
\bibitem{08-mixed} E. Jung, M. R. Hwang, D. K. Park, J. W. Son and S. Tamaryan,
{\it Mixed-state entanglement and quantum teleportation through noisy channels},
J. Phys. A: Math. Theor. {\bf 41} (2008) 385302 [arXiv:0804.4595 (quant-ph)].
\bibitem{karl98} A. Karlsson and M. Bourennane, {\it Quantum teleportation using
three-particle entanglement}, Phys. Rev. {\bf A58} (1998) 4394.
\end{thebibliography}
\end{document}